\documentclass[10pt]{iopart}
\usepackage{epsfig}
\usepackage{graphicx}
\newcommand{\be}{\begin{eqnarray}}
\newcommand{\ee}{\end{eqnarray}}

\newcommand{\ave}[1]{\left\langle #1 \right\rangle}

\begin{document}
\title{Phenomenology of Strangeness enhancement in heavy ion collisions}
\author{Giorgio Torrieri}
\address{Institut f\"ur Theoretische Physik,
  J.W. Goethe Universit\"at, Frankfurt A.M., Germany}

\date{September, 2007}

\begin{abstract}
We give an overview of the phenomenology of strangeness enhancement in heavy ion collisions, within the paradigm of the statistical model of particle production. We argue that, while strangeness enhancement data is suggestive of a phase transition, the mechanism triggering enhancement is still elusive.
We study the feasibility to constrain this mechanism by determining the scaling variable of strangeness enhancement.
We further argue that to test the applicability of the statistical model generally, and to confirm our interpretation of the statistical physics responsible for strangeness enhancement, the scaling of strange particle fluctuations ($K/\pi$ and other particles) w.r.t. yields has to be explored.
\end{abstract}
\maketitle
\section{Theoretical motivation}

The objective of heavy ion physics is to produce a  locally equilibrated state of quarks and gluons, the ``quark-gluon plasma'' (QGP), and to study the properties of this new phase, as well as the transition between quark-gluon plasma and hadronic matter.

One of the first proposed signatures for the new state of matter was strangeness enhancement \cite{raforig,raforig2}.   The basic idea was to look at the abundance of strange and (especially) multi-strange particles in A-A collisions, and compare with the corresponding $p-p$ and $p-A$ collisions (all scaled by the number of participants $N_{part}$ or the multiplicity $dN/d\eta$).  The appropriate observable is the ``enhancement'' $E$ of strange particle $Y$
\begin{equation}
E_{Y} = \frac{\ave{N_{part}^{p-A}}}{\ave{N_{part}^{A-A}}} \frac{\ave{N_Y^{A-A}}}{\ave{N_Y^{p-A}}}\phantom{AA}\mathrm{or}\phantom{AA}E_{Y} = \frac{\left( dN/d\eta \right)_{p-A}}{\left(dN/d\eta \right)_{A-A}} \frac{\ave{N_Y^{A-A}}}{\ave{N_Y^{p-A}}}
\label{enhdef}
\end{equation}
(We shall comment on the two definitions later.  Approximately, they give the same result, through the corrections could significantly modify the scaling).

It is expected to that $E_{Y}>1$ and increasing with strangeness content of the particle.
The reason given in \cite{raforig} is that strangeness-producing processes in a QGP
\begin{equation}
q\phantom{A}\overline{q} \leftrightarrow s \phantom{A} \overline{s}\phantom{AAAAAA}g \phantom{A}g \leftrightarrow s \phantom{A} \overline{s}
\label{procqgp}
\end{equation}
should equilibrate faster than the corresponding processes in a hadron gas (HG)
\begin{equation}
\pi^+ \phantom{A} \pi^- \leftrightarrow K\phantom{A} K \phantom{AAAAAA} \pi \phantom{A} N \leftrightarrow \Lambda K 
\label{prochg}
\end{equation}
this can be seen relatively quickly by computing the momentum exchange $\ave{Q}$ needed for these processes ($\ave{Q} \sim 2 m_s$ for Eq. \ref{procqgp},$\ave{Q} \sim m_K$ for Eq. \ref{prochg}. In a thermally equilibrated medium the equilibration time depends on $\ave{Q}/T$).  The greater degeneracy of massless quarks and gluons with respect to pions and nucleons makes the difference of thermalization times for the two phases even larger.
Hence, the strangeness abundance should reach chemical equilibrium (where the strangeness relative to light quark abundance depends only on mass difference and temperature, not on initial conditions)  much faster in a quark gluon plasma than in a hadron gas.    Since the initial strangeness abundance in collisions is zero, the number of strange particles in a system of a certain lifetime {\em with} a phase transition should be parametrically higher than for a similar system where the transition did not occur.  

This is particularly true for {\em multi-} strange particles: In the hadron gas phase, these can only be produced {\em sequentially}
\begin{equation}
\pi \phantom{A} \pi \phantom{A}\pi \phantom{A}N \leftrightarrow \pi \phantom{A}\pi \phantom{A} \Lambda \phantom{A} K  \leftrightarrow \pi \phantom{A} \Xi \phantom{A} K \phantom{A} K \leftrightarrow \Omega \phantom{A}K\phantom{A}K\phantom{A}K
\end{equation}
so the equilibration time for these is parametrically $N_{s} \tau_\Lambda$.  In a hadronizing quark-gluon plasma, strange hadrons are presumably produced via {\em coalescence} of quarks.  Hence, $\Lambda,\Xi$ and $\Omega$ are ``automatically'' equilibrated to the same degree.

In summary, strangeness particle abundance in a collision where a QGP is produced (a high energy A-A collision) should be enhanced w.r.t. a collision where hadronic dynamics is at play (a p-p or p-A collision, or an A-A collision where hadronic dynamics dominates).  The enhancement should also grow with the hadron's strangeness content.
The onset of this enhancement could signal the appearance of a phase transition, or more generally a change in the degrees of freedom of the system.

Experimental data have soundly confirmed this prediction \cite{strangesps,strangesps2,strangerhic}.  Both at SPS and RHIC, enhancement defined w.r.t. p-p collisions is unity for p-A and d-A.   At A-A, it increases with both $N_{part}$ and strangeness content of the particle, $\ln E \propto s \ln N_{part}$.   The constant of proportionality does not vary between RHIC and SPS. 
This is exactly the behavior expected under the hypothesis that, in A-A (but not in p-A) systems the {\em density} of strangeness is higher than in p-p.   

Given that no evidence of such a scaling violation has been found in other soft heavy ion observables ($v_2$ \cite{v2scaling} 
and HBT radii \cite{lisa} scale remarkably smoothly with multiplcitiy), and the difficulty of doing scaling studies with Hard 
observables, it is worth underlying that the scaling behavior of strangeness enhancement is {\em very} reminiscent of the 
behavior associated with a phase transition.   {\em If} we find an energy where enhancement of multi-strange particles does 
{\em not} occur (the coming low-energy run at FAIR \cite{fair,fair2} and the SHINE program at the SPS \cite{shine} will be an 
optimal opportunity to look for such 
a 
system), we would conclusively confirm that strangeness is an ``order parameter'' where scaling violation occurs for a critical energy and system size.   

However, the interpretation of this observable has {\em not} been universally agreed upon.  In particular, it was pointed out that the same behavior of enhancement w.r.t. strangeness can be obtained through strictly conserving strangeness \cite{canonical1}.   Recently, a change in the thermalization time through the excitation of higher {\em hadronic} resonances and/or multi-particle processes \cite{jaki} has been suggested to produce the same effects.

The focus of this work is to review these hypotheses,in particular  how strangeness enhancement {\em scales} with particle type, energy, system size in each.

\section{Statistical hadronization phenomenology of strangeness enhancement}
The idea of modelling the abundance of hadrons using statistical mechanics techniques has a long and distinguished history ~\cite{Fer50,Pom51,Lan53,Hag65}.  In a sense, any discussion of the thermodynamic properties of hadronic matter (e.g. the existence of a phase transition) {\em requires} that statistical mechanics be applicable to this system ( through not necessarily at the freeze-out stage).
That such a model can in fact describe {\em quantitatively} the yield of most particles, including multi-strange ones, has in fact been indicated by fits to average particle abundances at AGS,SPS and RHIC energies  ~\cite{bdm,equil_energy,jansbook,becattini,nuxu,share,sharev2}.
We refer to some of these references for a review of the statistical model.    For this work, it is sufficient to say that the particle abundance can be calculated using the following formula:
\begin{equation}
\ave{N_i} = f\left( V,Q-\overline{Q} \right) \prod_i \left( \lambda_q^{Q_i - \overline{Q}_i} \gamma_q ^{Q_i + \overline{Q}_i} \right) F \left( \frac{m}{T} \right) + \sum_{j \rightarrow i} b_{j
\rightarrow i} \ave{N_j}
\label{statmod}
\end{equation}
we now examine each term, explaining its physical origin and its scaling properties.

The resonance feed-down contribution, $\sum_{j \rightarrow i} b_{j
\rightarrow i} \ave{N_j}$ (in terms of the Branching ratio $b_{j \rightarrow i}$ and the resonance abundance $\ave{N_j}$) is straight-forward to take into account, {\em provided} the full table of resonances is available.   In the calculations of this work, we have used the data-base in SHARE \cite{share,sharev2}.

The thermo-statistical weight, in both the Canonical and Grand-Canonical ensembles, $ F \left( \frac{m}{T} \right)$, is approximately given in terms modified Bessel function \cite{jansbook} of the ratio of the mass of the particle $m$ as well as the chemical freeze-out temperature.   Since strange particles are heavier than light particles, a higher/lower chemical freeze-out temperature could give rise to an effective strangeness enhancement/suppression. 
It is worth underlining that freeze-out temperature is highly correlated, as a fit parameter, to the phase space factors $\gamma_Q$ examined later \cite{share,sharev2}.

The Volume factor in Eq. \ref{statmod}, $f\left( V,Q-\overline{Q} \right)$, is trivially proportional to the volume, $\sim V$ in the thermodynamic limit (the limit where $V/V_0 \gg 1$, where $V_0$ is the ``typical scale'' of one unit of charge $Q$) and in the Grand-Canonical ensemble.  
If strangeness abundance is really described in this limit, then {\em all} volume effects from the strangeness enhancement can be eliminated by defining the enhancement in terms of multiplicity (right hand definition of Eq. \ref{enhdef}), since multiplicity, in the Grand Canonical ensemble, tracks the system volume at freeze-out \footnote{$N_{part}$ tracks the system volume at the start of the evolution.  Deviations from isothropic expansion, initial state effects, and event-by-event fluctuations of the dynamics \cite{v2fluct} could well make the relationship between initial and final volume non-trivial, thereby spoiling any scaling in the strangeness enhancement.  Because of these ambiguities, we favour defining $E_Y$ in terms of the multiplicity, through the definition in terms of $N_{part}$ is more widely used} 

Away from the thermodynamic limit ($V/V_0 \leq 1$), it becomes ensemble-specific. 
In the Canonical and Micro-Canonical ensembles, it will scale, approximately, as the volume to the power of the total {\em  net} charge $Q$ of the particle, $F(Q,V) \sim V^{|Q-\overline{Q}|}$ \cite{canonical1}, since the phase space {\em exactly} conserving the charge of a small system producing a particle of charge $Q$ is (approximately exponentially with $Q$) suppressed w.r.t. the thermodynamic limit expectation. In the Grand canonical ensemble charge violation is allowed event by event, due to the exchange of charge between the system and the bath.   Thus, the particle yield is proportional to volume for all volumes.

Because of this effect, if the smaller system is away from the thermodynamic limit, the larger system is closer to it {\em and} charge is conserved for the observable system \footnote{Not guaranteed at RHIC experiments, where charge could well be exchanged between the observed central rapidity region and the ``bath'' given by the rest of the system. The non-observation of neutral particles in most experiments also spoils observable effects of conservation laws} the Enhancement calculated according to Eq \ref{enhdef} is
\begin{equation}
E_{Y} \sim \left( f \left[ \frac{N_{part}^{AA}}{ N_{part}^{p-A}} \right] \right)^{\left| Q_{Y} - \overline{Q}_{Y} \right|},f(x \ll 1)\propto x\phantom{A},\lim_{x\rightarrow \infty} f(x)=x^{MAX}\phantom{AA}{\mathrm(Constant)}\phantom{AAA}
\end{equation}
 This behavior, and the scaling w.r.t. $Q-\overline{Q}$ are {\em experimental signatures} capable of ascertaining the contribution of canonical suppression to strangeness enhancement.

The chemical factor in Eq. \ref{statmod}, $\prod \left( \lambda_Q^{Q_i - \overline{Q}_i} \gamma_Q^{Q_i + \overline{Q}_i} \right)$, counts the contributions of the {\em density} of the conserved charge inside the system, parameterized by the fugacity $\lambda_Q$ and the phase space factor $\gamma_Q$.   The first factor, $\lambda_Q$, counts the net charge density of the system, can not give rise to strangeness enhancement (since the net strange density is zero independently of strangeness density) and, as it commutes with the Hamiltonian, it is independent of the degree of chemical equilibration of the system.
To account for chemical non-equilibrium, another parameter, $\gamma_Q$, can be introduced.  This counts the abundance of $Q \overline{Q}$ {\em pairs} in the system.   It is normalized to be $=1$ at equilibrium and in general, does not commute with the Hamiltonian.

If strangeness equilibration indeed proceeds faster in the QGP phase, then $\gamma_s$ is expected to be $<1$ in p-A systems and $\simeq 1$ in A-A systems.
It is then trivial to show that enhancement scales as $\gamma_s^{s+\overline{s}}$.
Note that the exponent in the enhancement factor is the {\em total} strangeness $s+\overline{s}$, rather than the {\em net} strangeness in the earlier canonical suppression scenario ($\gamma_s$ increases the {\em density} of strange particles). For most particles, these two numbers are the same, but not for all.  For instance, in $\phi$ $s-\overline{s}=0$, but $s+\overline{s}=2$.    Hence, in the canonical enhancement scenario the $\phi$ should not be enhanced at all, but in the scenario where chemical equilibration dominates, it should be {\em as} enhanced as the $\Xi$.   Other mesons with an $s\overline{s}$ content, such as the $\eta$, should behave accordingly (SHARE \cite{share} takes the mixing of quarks into account in fits including $\gamma$).  Thus, the two scenarios are {\em not} different interpretations, but different theories, falsifiable by an accurate study of the scaling behavior of enhancement.

The dependence of strangeness enhancement w.r.t. $N_{part}$ could however be similar in both cases: Strangeness enhancement, as we have seen, saturates in the large $N_{part}$ limit since the system there reaches the thermodynamic limit.
In the ``faster equilibration in QGP'' scenario, $\gamma_s$ should also approach the saturating value $1$ asymptotically {\em provided} the system obeys a kinetic type evolution equation {\em throughout} the phase transition and the hadronic phase.  In this case, the enhancement is
\begin{equation}
E_{Y} \sim \left( \frac{\gamma_s^{A-A}}{\gamma_s^{p-A}} \right)^{s+\overline{s}}\phantom{A},\phantom{A}\lim_{N_{part}\rightarrow\infty} \gamma_s^{A-A} =1
\end{equation}
However, there is (possibly) more to this issue:  the {\em equilibrium} strange particle abundance in an {\em ideal} QGP should be parametrically higher than the corresponding equilibrium abundance of strange quarks in a HG 
\begin{equation}  
\left. \frac{n_s}{n_q} \right|_{QGP} \sim e^{\frac{m_s - m_q}{T}},
\left.
\frac{n_s}{n_q}\right|_{HG}^{\lambda_Q\simeq 1} \sim e^{\frac{m_K - m_{\pi}}{T}},\left. \frac{n_s}{n_q}\right|_{HG}^{\lambda_Q\gg1}\sim e^{\frac{m_\Lambda - m_{p}}{T}}
\end{equation} 
Hence, the abundance of strange particles at $T>T_c$ should be {\em above} that for a perfectly equilibrated hadron gas.
How  does such an abundance evolve when the basic degrees of freedom of the system change?   Dimensional analysis indicates that this can be understood via the dimensionless parameter 
\begin{equation}
\alpha = \left[\frac{\chi_{s}(T,\mu_B)}{\rho_s(T,\mu_B)}\right]\left[\frac{d T}{d \tau} + \frac{d \mu_B}{d \tau}\right] \tau_{s}
\label{alphaeq}
\end{equation}
where $\chi_s,\rho_s$ are the strangeness susceptibility and equilibrium density in the hot phase, $\frac{d T}{d \tau}$ and $\frac{d \mu_B}{d \tau}$ are the rates of change in temperature and baryochemical potential in the co-moving frame and $\tau_{s}(T,\mu_B)$ is the equilibration time-scale for $s$ close to the phase transition.
If $\alpha \ll 1$, the system's chemical composition changes smoothly from the old phase to the new phase equilibrium value as the system undergoes the transition.   If $\alpha \gg 1$, the system does not have time to adjust its chemical composition, and the colder phase is produced {\em out} of chemical equilibrium, reflecting the degree of chemical equilibrium of the {\em hot} phase.  In this case, it is expected that $\gamma_s>1$, a value {\em forbidden} in a Boltzmann equation (and in general unlikely unless a fast change in microscopic degrees of freedom occurs)

For the case of strangeness, $\chi_s/\rho_s$ is known from the lattice to have a peak, but not a divergence (in accordance to the current understanding of deconfinement to be a cross-over, not a first order phase transition).
$\frac{d T}{d \tau}$ can be estimated through hydrodynamics.
$\tau_{s}$ is completely unknown.   The chemical equilibration timescale 
is, however, thought to be related to the bulk viscosity, which is 
presumed to exhibit a strong peak \cite{bulk1,bulk2,bulk3}.  Thus, 
$\alpha$ 
might well be considerably larger than a lattice estimate, together with a naive extrapolation (e.g. $\tau_s \sim 1/T$) might suggest.   

The {\em exact same} reasoning, however, holds for the {\em light quark} degrees of freedom $q$, which could therefore be similarly {\em oversaturated} with respect to equilibrium (most statistical models assume them to be equilibrated).  If this is the case, both $\gamma_q$ and $\gamma_s$ should be used, and are expected to be $>1$ at freeze-out (their exact value is not readily calculable, through it can be estimated by entropy matching \cite{jansbook})

In this scenario, several factors contribute to strangeness enhancement: One is greater strangeness content (due to $\gamma_s>1$).  Another is the number of quarks ($\gamma_q>1$ so Baryons are more enhanced than mesons).  The onset of the saturation $\gamma_{q,s}\rightarrow 1$ for a long-lived ($N_{part}$ large) system need not occur

Putting all these points together, while the interpretation of strangeness enhancement is not univocal, different models predict different {\em scaling variables} for the enhancement.   If strangeness enhancement is due to canonical suppression, it should increase with {\em net} strangeness $|s-\overline{s}|$ of the hadron.   If it is due to a growth of strangeness phase space density, it should   grow with {\em total} strangeness $s+\overline{s}$ of the hadron.   If the phase space density saturates to its equilibrium value up to freezeout, strangeness enhancement should also saturate.  If, however, lack of equilibrium is due to a fast transition from a phase with different equilibrium values for strange and light quark density, this saturation does not need to occur.  In this latter scenario, the strangeness enhancement need not depend entirely on strangeness content, but can also show a Baryon/Meson dependence due to light quark non-equilibrium.   

A change with freeze-out temperature with system size (expected if the chemical non-equilibrium is associated with supercooling, as in \cite{jansbook}) could give a further scaling of enhancement with particle mass.  A non-linear dependence of the final freeze-out volume with $N_{part}$ (expected if expansion is not quite isentropic) could lead to an enhancement contribution for {\em all} particles, independently of mass and chemical content (Defining $E_Y$ with multiplicity would eliminate this effect).   A reinteracting hadron gas stage after hadronization could affect {\em resonance enhancement} in a non-trivial way \cite{kuzraf} (through this effect is expected to be $\sim \Gamma_i\tau_{HG}$,where $\Gamma_i$ is the width of the resonance and $\tau_{HG}$ the lifetime of the interacting HG). Since several changes could occur at the same time, inverting the observed enhancements to obtain what statistical parameter changes between p-A and A-A systems is non-trivial.

\begin{figure}[h]
\begin{center}
\epsfig{width=13cm,figure=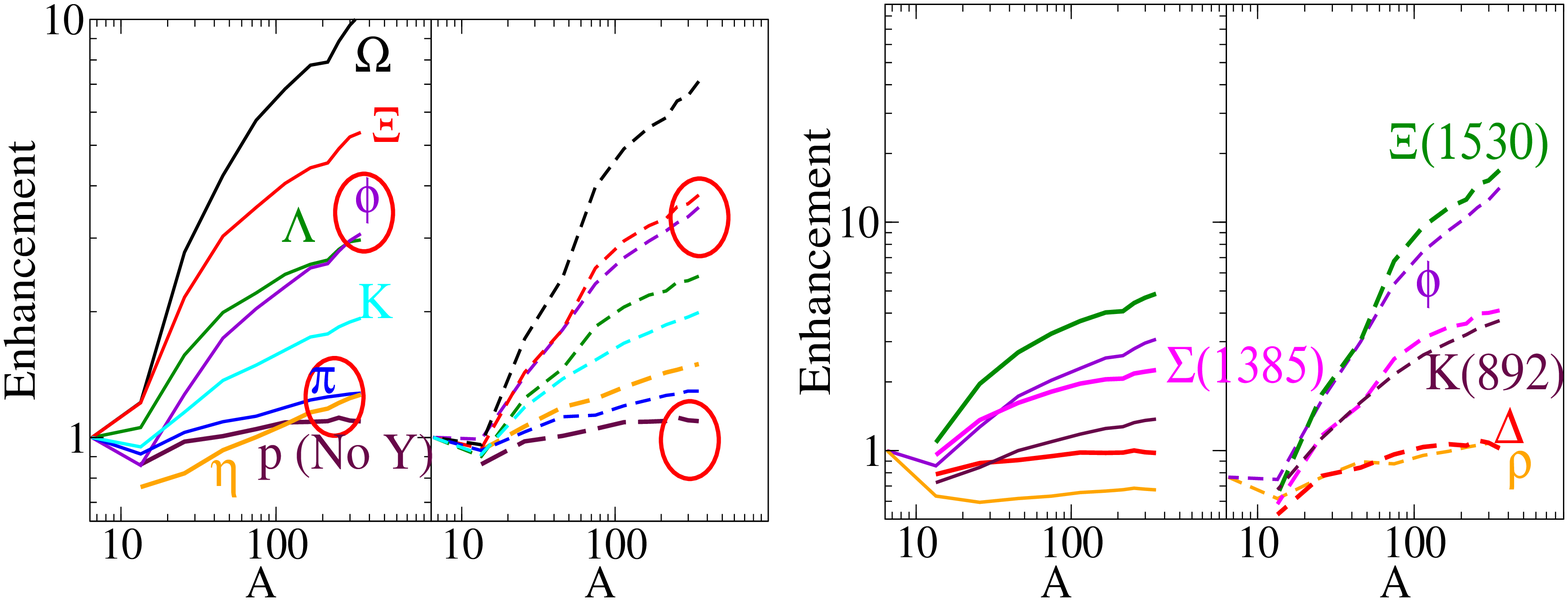}
\caption{\label{enhfig} (Color online) Enhancement factors of different stable particles (left panel) and resonances (right panel) for the equilibrium (dashed lines) and non-equilibrium (solid lines) statistical models.  Red ellipses show where the scaling in the models differs }
\end{center}
\end{figure}

An explicit calculation of enhancements from both the equilibrium and non-equilibrium statistical model (parameter fits taken from \cite{ourrhicfit}, relevant for RHIC energies) is shown in Fig. \ref{enhfig}
Summarizing this section, different scenarios for the origin of strangeness enhancement within the statistical model give, generally, different scaling observables controlling strangeness enhancement.  The numerous competing effects, however, render it non-trivial to conclusively establish which model is more physically appropriate by scaling studies {\em alone}.  The applicability of statistical mechanics at all energies and system sizes also remains untested.  We now propose that the scaling of strange particle {\em fluctuations} is necessary to address these points.
\section{Scaling of fluctuations as a confirmation of statistical behavior}
Particle fluctuations are a promising observable to falsify the statistical model and to constrain its parameters (choice of ensemble, strangeness/light quark chemical equilibrium) \cite{prcfluct}.
One can immediately see that fluctuations are a stringent statistical model test by considering the fluctuation of a ratio between two random variables.
\begin{equation}
\label{fluctratio}
\sigma_{N_1/N_2}^2
= \frac{\ave{(\Delta N_1)^2}}{\ave{N_1}^2}
+ \frac{\ave{(\Delta N_2)^2}}{\ave{N_2}^2}
- 2 \frac{\ave{\Delta N_1 \Delta N_2}}{\ave{N_1}\ave{ N_2}}.
\end{equation}
Since, for an equilibrated system, in the thermodynamic limit $\ave{(\Delta N_1)^2} \sim \ave{N} \sim \ave{V}$, where $\ave{V}$ is the system volume,
it is clear that $\sigma_{N_1/N_2}^2$ depends on the hadronization volume in a manner {\em opposite} to that of particle yields.    Volume fluctuations (which make a comparison of statistical model calculations to experimental data problematic), both resulting from dynamics and from fluctuations in collision geometry, should not alter this very basic result since volume cancels out of particle ratios event by event \cite{jeon}, {\em provided } hadronization volume is the same for all particles (a basic statistical model requirement).

  Thus, observables such as $\ave{N_1} \sigma_{N_1/N_2}^2$,
provided $\ave{N_{1,2}}$ and $\sigma_{N_1/N_2}^2$ are measured using the same kinematic cuts, 
 should be  strictly independent of multiplicity and centrality, as long as the statistical model holds.  Furthermore, the {\em value} of this constant should be calculable with the same parameters used to calculate particle yields ($T,\lambda_{q,s},\gamma_{q,s}$ and {\em choice of ensemble} \cite{gazdzicki})  This is a stringent test, because variables which generally correlate for yields (such as $T$ and $\gamma_{q,s}$) {\em anti}-correlate for fluctuations \cite{prcfluct,sqm2006}.

The independence of $\ave{N_1} \sigma_{N_1/N_2}^2$ with volume can be stringently tested: {\em if} the temperature and chemical potentials between two energy regimes are approximately the same at freezeout (this should be the case for RHIC top energies and LHC, {\em provided chemical equilibrium holds}), $\ave{N_1} \sigma_{N_1/N_2}^2$ should also stay constant across energies.  $\ave{N_1} \sigma_{N_1/N_2}^2$ should also be approximately the same for Au-Au and Cu-Cu collisions at high RHIC energy \cite{sqm2006}, and independent of centrality.
  If $\gamma_{q,s}$ at SPS,RHIC and LHC are oversaturated w.r.t. equilibrium (As in the scenario described in \cite{jansbook}), then $\ave{N_1} \sigma_{N_1/N_2}^2$ should still be independent of centrality and system size (Cu-Cu vs Au-Au) for a given energy range, but should go markedly up for the LHC from RHIC, because of the increase in $\gamma_q$ and $\gamma_s$.  

While fluctuations are more sensitive to acceptance cuts then yields, these systematic errors can be at least partially removed via mixed event subtraction \cite{methods}.
An appropriate observable to model and measure  \cite{methods,supriya,spsfluct} is therefore $\sigma_{dyn}^2 = \sigma^2 - \sigma^2_{mix}$
where  $\sigma^2_{mix}$ is the mixed event width\footnote{the $\nu_{dyn}$ observable, equivalent to $\sigma_{dyn}^2$ for large systems \cite{methods}, is more appropriate for rarer particles}.    In the absence of any correlations, $\sigma_{mix}$ reduces to the Poisson expectation, $\sigma_{mix}^2 = 1/\ave{N_1}+1/\ave{N_2}$. We therefore propose to use 
$\Psi^{N_1}_{N_1/N_2} = \ave{N_1} \sigma_{dyn}^{N_1/N_2}$
to test the statistical model among different energy, system size and centralities.  
 
\begin{figure}[h]
\begin{center}
\epsfig{width=13cm,figure=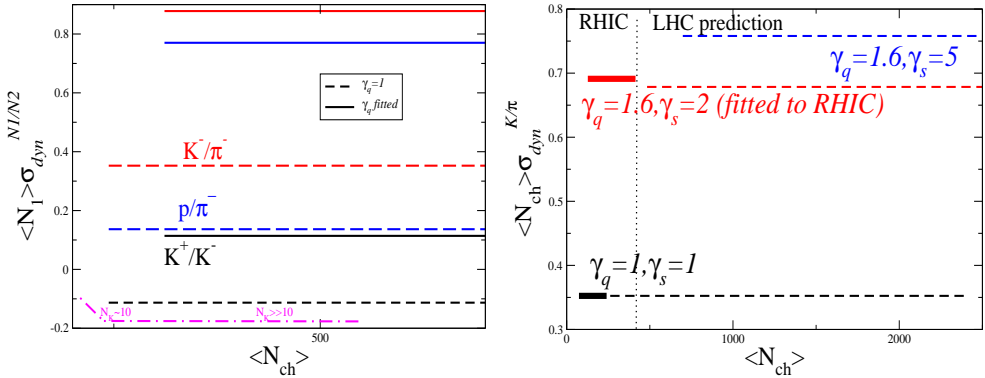}
\caption{\label{figfluct}  Scaling of the $\Psi$ observable w.r.t. multiplicity for particle ratios at RHIC (left panel) and LHC (right panel) energies in the $\gamma_{q,s}>1$ non-equilibrium scenario, equilibrium scenario and equilibrium-canonical scenario}
\end{center}
\end{figure}

We have calculated $\Psi^{N_1}_{N_1/N_2}$ for RHIC and LHC energies, for the sets of parameters used in \cite{janlhc}.  The left and right panel in  Fig. 
\ref{figfluct} shows what effect three different sets of $\gamma_{q,s}$ inferred in \cite{janlhc} would have on  $\Psi^{\pi^-}_{K^-/\pi^-}$ and $\Psi^{\pi^-}_{K^-/K^+}$.   In the left panel we have also included the value of  $\Psi^{\pi^-}_{K^-/\pi^-}$ for top energy RHIC.   As shown in \cite{sqm2006}, this value for top centrality matches expectations for non-equilibrium freeze-out (and is significantly above equilibrium freeze-out).   A centrality scan of  $\Psi^{\pi^-}_{K^-/\pi^-}$, necessary to confirm the consistency of this result has not, however, as yet been performed.

The scaling should break if global correlations persist. such as is the case if the Canonical and micro-canonical ensembles \cite{gazdzicki} are physically more appropriate to describe the system (and Canonical suppression contributes to strangeness enhancement).  If strangeness at RHIC/the LHC is created and maintained locally,  $\Psi^{N_1}_{N_1/N_2}$ should develop a ``wiggle'' at low centrality, and be lower (by half for $\Psi^{\pi^-}_{K^+/K^-}$) than Grand Canonical expectation. 

In conclusion, we have outlined the different statistical models capable of explaining the experimentally observed onset of strangeness enhancement, and argued that the scaling of enhancement with hadron type is generally different in each of these models.  We have further argued that {\em all} statistical models can be stringently tested by observing the scaling of strange particle fluctuations w.r.t. yields.   

G.T. acknowledges the financial support received from the Helmholtz International
Center for FAIR within the framework of the LOEWE program
(Landesoffensive zur Entwicklung Wissenschaftlich-\"Okonomischer
Exzellenz) launched by the State of Hesse.
We also thank the SQM2008 organizing committee for  
support, and J.Rafelski, M.Hauer and A. Timmins for useful discussions

\end{document}